\documentclass[10pt, conference]{IEEEtran}
\usepackage{amsmath,epsfig}
\usepackage[applemac]{inputenc}
\usepackage{amssymb}
\usepackage{amsthm}
\usepackage{color, colortbl}
\usepackage{booktabs}
\usepackage{multirow}
\usepackage[cmintegrals]{newtxmath}

\usepackage{hhline}
\usepackage{multicol,float}
\usepackage{pgfplots}
\usetikzlibrary{plotmarks}
\usepackage{tabularx}
\usepackage{tikz}
\usepackage{verbatim}
\usepackage{stmaryrd} 
\usepackage{epstopdf}
\usepackage{epsfig}
\usepackage{tkz-berge}
\usepackage{balance}

\def\bds#1{\boldsymbol{#1}}

\def\bdm#1{\mathbf{#1}}

\def\tildeC{\widetilde{\bdm{C}}}

\def\Bbb#1{\mbox{I\hspace{-.15em}#1}}

\graphicspath{/figures}

\newtheorem*{theorem}{Consistency theorem}

\title{Improving Portfolios Global Performance with Robust Covariance Matrix Estimation: \\Application to the Maximum Variety Portfolio}
\date{}
\author{\IEEEauthorblockN{Emmanuelle Jay\IEEEauthorrefmark{1}\IEEEauthorrefmark{2}, Eug\'enie Terreaux\IEEEauthorrefmark{4}, Jean-Philippe Ovarlez\IEEEauthorrefmark{3} and Fr\'ed\'eric Pascal\IEEEauthorrefmark{5}} \IEEEauthorblockA{\IEEEauthorrefmark{1}Fideas Capital, 
21 avenue de l'Op\'era, 75001 Paris, France - Email: ejay@fideas.fr} \IEEEauthorblockA{\IEEEauthorrefmark{2}Quanted \& Europlace Institute of Finance, Palais Brongniart, 28 place de la Bourse, 75002 Paris, France}
\IEEEauthorblockA{\IEEEauthorrefmark{3}ONERA, DEMR/TSI, Chemin de la Huni\`ere, 91120 Palaiseau, France}
\IEEEauthorblockA{\IEEEauthorrefmark{4}CentraleSup\'elec, 
3 rue Joliot-Curie, 91190 Gif-sur-Yvette, France} \IEEEauthorblockA{\IEEEauthorrefmark{5}L2S / CentraleSup\'elec - CNRS - Universit\'e Paris-Sud, 
3 rue Joliot-Curie, 91190 Gif-sur-Yvette, France}
}

\begin{document}

\maketitle

\begin{abstract}
This paper presents how the most recent improvements made on covariance matrix estimation and model order selection can be applied to the portfolio optimisation problem. The particular case of the Maximum Variety Portfolio is treated but the same improvements apply also in the other optimisation problems such as the Minimum Variance Portfolio. We assume that the most important information (or the latent factors) are embedded in correlated Elliptical Symmetric noise extending classical Gaussian assumptions. We propose here to  focus on a recent method of model order selection allowing to efficiently estimate the subspace of main factors describing the market. This non-standard model order selection problem is solved through Random Matrix Theory and robust covariance matrix estimation. The proposed procedure will be explained through synthetic data and be applied and compared with standard techniques on real market data showing promising improvements. 
\end{abstract}


\section{Introduction}

Portfolio allocation is often associated with the mean-variance framework fathered by Markowitz in the 50's \cite{markowitz52}. This framework designs the allocation process as an optimisation problem where the portfolio weights are such that the expected return of the portfolio is maximised for a given level of portfolio risk. In practice this needs to estimate both expected returns and covariance matrix leading to estimation errors, particularly important for expected returns. 
This partly explains why many studies concentrate on allocation process relying solely on the covariance estimation such as the Global Minimum Variance Portfolio or the Equally Risk Contribution Portfolio \cite{clarke12}, \cite{maillard10}.
Another way to reduce the overall risk of a portfolio is to diversify the risks of its assets and to look for the assets weights that maximise a diversification indicator such as the variety (or diversification) ratio \cite{choueifaty08,choueifaty13}, only involving the covariance matrix of the assets returns as well.\\
\indent The frequently used covariance estimator is the Sample Covariance Matrix (SCM), optimal under the Normal assumption. Financial time series of returns might exhibit outliers related to abnormal returns leading to estimation errors larger than expected. 
The field of robust estimation \cite{Tyler87}, \cite{maronna76} intends to deal with this problem especially when $N$, the number of samples, is larger than $m$, the size of the observations vector. When $N<m$, the covariance matrix estimate is not invertible and regularization approaches are required. Some authors have proposed hybrid robust shrinkage covariance matrix estimates \cite{chen11}, \cite{pascal14}, \cite{abramovich07}, building estimators upon Tyler’s robust M-estimator \cite{Tyler87} and Ledoit-Wolf’s shrinkage approach \cite{ledoit04}.\\

\indent Recent works \cite{chen11}, \cite{couillet14a}, \cite{pascal14}, \cite{yang15} based on Random Matrix Theory (RMT) have therefore considered robust estimation in the $m,N$ regime.  
In \cite{yang15}, the Global Minimum Variance Portfolio is studied and the authors show that applying an adapted estimation methodology based on the Shrinkage-Tyler M-estimator leads to achieving superior performance over may other competing methods.
Another way to mitigate covariance matrix estimation errors is to filter the noisy part of the data. In financial applications, several empirical evidence militate in favour of the existence of multiple sources of risks challenging the CAPM single market factor assumption \cite{sharpe64}. Whereas statistical methods like the principal component analysis may fail in distinguishing informative factors from the noisy ones, 
RMT helps in finding a solution for filtering noise \cite{laloux99, laloux00, potters05, plerou01},
even though the single market factor still prevails in the described cleaning method that is not completely satisfactory.\\
\indent The application here proposes to mix several approaches: the assets returns are modelled as a multi-factor model embedded in correlated elliptical and symmetric noise and the final covariance estimate will be computed on the "signal only" part of the observations, separable from the "noise part" thanks to the results found in \cite{Vinogradova13, terreaux17, terreaux17a, terreaux18}.\\
\indent The article is constructed as follows: section II presents the classical model and assumptions under consideration. Section III introduces the selected method of portfolio allocation for this paper: the Maximum Variety portfolio. Section IV explains how to solve the problem jointly with RMT and robust estimation theory which allow to design a consistent estimate of the number $K$ of informative factors. Section V shows some results obtained on experimental financial data highlighting the efficiency of the proposed method with regards to the conventional ones. Conclusion in section VI closes this paper. 
\begin{figure*}
\begin{tabular}{ccc}
%
%
\definecolor{mycolor1}{rgb}{0.00000,0.44700,0.74100}%
\begin{tikzpicture}

\begin{axis}[%
width=0.5\columnwidth,
at={(1.011in,0.642in)},
scale only axis,
bar shift auto,
xmin=-2.45,
xmax=4.96,
xlabel style={font=\color{white!15!black}},
xlabel={$\log{\lambda_i}$},
ymin=0,
ymax=1200,
axis background/.style={fill=white},
title style={font=\bfseries},
xmajorgrids,
ymajorgrids,
legend style={legend cell align=left, align=left, draw=white!15!black}
]
\addplot[ybar interval, fill=mycolor1, fill opacity=0.6, draw=black, area legend] table[row sep=crcr] {%
x	y\\
-1.65	2\\
-1.5919	13\\
-1.5338	74\\
-1.4757	228\\
-1.4176	389\\
-1.3595	586\\
-1.3014	662\\
-1.2433	760\\
-1.1852	881\\
-1.1271	868\\
-1.069	978\\
-1.0109	1037\\
-0.9528	1031\\
-0.8947	1044\\
-0.8366	1077\\
-0.7785	1106\\
-0.7204	1110\\
-0.6623	1152\\
-0.6042	1090\\
-0.5461	1077\\
-0.488	1098\\
-0.4299	1070\\
-0.3718	988\\
-0.3137	1023\\
-0.2556	1006\\
-0.1975	957\\
-0.1394	959\\
-0.0812999999999999	889\\
-0.0231999999999999	835\\
0.0348999999999999	848\\
0.093	829\\
0.1511	773\\
0.2092	768\\
0.2673	699\\
0.3254	667\\
0.3835	673\\
0.4416	652\\
0.4997	631\\
0.5578	614\\
0.6159	558\\
0.674	573\\
0.7321	565\\
0.7902	556\\
0.8483	578\\
0.9064	595\\
0.9645	525\\
1.0226	509\\
1.0807	467\\
1.1388	423\\
1.1969	459\\
1.255	438\\
1.3131	408\\
1.3712	439\\
1.4293	454\\
1.4874	444\\
1.5455	422\\
1.6036	362\\
1.6617	376\\
1.7198	328\\
1.7779	344\\
1.836	251\\
1.8941	334\\
1.9522	277\\
2.0103	318\\
2.0684	277\\
2.1265	259\\
2.1846	265\\
2.2427	272\\
2.3008	252\\
2.3589	261\\
2.417	262\\
2.4751	253\\
2.5332	214\\
2.5913	223\\
2.6494	260\\
2.7075	220\\
2.7656	231\\
2.8237	221\\
2.8818	229\\
2.9399	222\\
2.998	233\\
3.0561	215\\
3.1142	219\\
3.1723	216\\
3.2304	218\\
3.2885	245\\
3.3466	217\\
3.4047	209\\
3.4628	262\\
3.5209	260\\
3.579	264\\
3.6371	287\\
3.6952	274\\
3.7533	261\\
3.8114	259\\
3.8695	207\\
3.9276	116\\
3.9857	48\\
4.0438	19\\
4.1019	3\\
4.16	3\\
};
\addlegendentry{ \footnotesize eigenvalues}

\addplot[ybar, bar width=1.5, fill=green, draw=black, area legend] table[row sep=crcr] {%
0.549539784816691	800\\
};
\addplot[forget plot, color=white!15!black] table[row sep=crcr] {%
-2.45	0\\
4.96	0\\
};
\addlegendentry{\footnotesize $\bar{\lambda}$ threshold}

\end{axis}
\end{tikzpicture}%
 & 
%
%
\definecolor{mycolor1}{rgb}{0.00000,0.44700,0.74100}%
\begin{tikzpicture}

\begin{axis}[%
width=0.5\columnwidth,
at={(1.011in,0.642in)},
scale only axis,
bar shift auto,
xmin=-4.37,
xmax=4.46,
xlabel style={font=\color{white!5!black}},
xlabel={$\log{\lambda_i}$},
ymin=0,
ymax=2000,
axis background/.style={fill=white},
title style={font=\bfseries},
title={},
xmajorgrids,
ymajorgrids,
legend style={legend cell align=left, align=left, draw=white!15!black}
]
\addplot[ybar interval, fill=mycolor1, fill opacity=0.6, draw=black, area legend] table[row sep=crcr] {%
x	y\\
-3.57	5\\
-3.4977	6\\
-3.4254	26\\
-3.3531	74\\
-3.2808	164\\
-3.2085	357\\
-3.1362	712\\
-3.0639	1064\\
-2.9916	1364\\
-2.9193	1630\\
-2.847	1756\\
-2.7747	1836\\
-2.7024	1924\\
-2.6301	1844\\
-2.5578	1778\\
-2.4855	1728\\
-2.4132	1660\\
-2.3409	1490\\
-2.2686	1466\\
-2.1963	1308\\
-2.124	1222\\
-2.0517	1147\\
-1.9794	1065\\
-1.9071	1006\\
-1.8348	954\\
-1.7625	880\\
-1.6902	864\\
-1.6179	804\\
-1.5456	754\\
-1.4733	729\\
-1.401	683\\
-1.3287	664\\
-1.2564	632\\
-1.1841	612\\
-1.1118	587\\
-1.0395	544\\
-0.9672	546\\
-0.8949	500\\
-0.8226	480\\
-0.7503	473\\
-0.678	459\\
-0.6057	468\\
-0.5334	414\\
-0.4611	394\\
-0.3888	403\\
-0.3165	387\\
-0.2442	377\\
-0.1719	370\\
-0.0996000000000001	359\\
-0.0273000000000003	314\\
0.0449999999999999	332\\
0.1173	304\\
0.1896	335\\
0.2619	301\\
0.3342	294\\
0.4065	291\\
0.4788	296\\
0.5511	260\\
0.6234	283\\
0.6957	288\\
0.768	260\\
0.8403	271\\
0.9126	262\\
0.9849	251\\
1.0572	290\\
1.1295	292\\
1.2018	254\\
1.2741	308\\
1.3464	307\\
1.4187	347\\
1.491	400\\
1.5633	389\\
1.6356	370\\
1.7079	329\\
1.7802	253\\
1.8525	124\\
1.9248	57\\
1.9971	36\\
2.0694	35\\
2.1417	22\\
2.214	43\\
2.2863	57\\
2.3586	77\\
2.4309	63\\
2.5032	65\\
2.5755	71\\
2.6478	48\\
2.7201	47\\
2.7924	50\\
2.8647	47\\
2.937	67\\
3.0093	77\\
3.0816	90\\
3.1539	62\\
3.2262	53\\
3.2985	31\\
3.3708	13\\
3.4431	7\\
3.5154	5\\
3.5877	3\\
3.66	3\\
};
\addlegendentry{ \footnotesize eigenvalues}

\addplot[ybar, bar width=1, fill=green, draw=black, area legend] table[row sep=crcr] {%
0.549539784816691	800\\
};
\addplot[forget plot, color=white!15!black] table[row sep=crcr] {%
-4.37	0\\
4.46	0\\
};
\addlegendentry{\footnotesize $\bar{\lambda}$ threshold}

\end{axis}
\end{tikzpicture}%
 & 
%
%
\definecolor{mycolor1}{rgb}{0.00000,0.44700,0.74100}%
\begin{tikzpicture}

\begin{axis}[%
width=0.5\columnwidth,
at={(0.711in,0.642in)},
scale only axis,
bar shift auto,
xmin=-4.09,
xmax=4.73,
xlabel style={font=\color{white!15!black}},
xlabel={$\log{\lambda_i}$},
ymin=0,
ymax=2000,
axis background/.style={fill=white},
title style={font=\bfseries},
title={},
xmajorgrids,
ymajorgrids,
legend style={legend cell align=left, align=left, draw=white!15!black}
]
\addplot[ybar interval, fill=mycolor1, fill opacity=0.6, draw=black, area legend] table[row sep=crcr] {%
x	y\\
-3.29	2\\
-3.2178	5\\
-3.1456	13\\
-3.0734	11\\
-3.0012	20\\
-2.929	31\\
-2.8568	51\\
-2.7846	75\\
-2.7124	127\\
-2.6402	202\\
-2.568	306\\
-2.4958	425\\
-2.4236	618\\
-2.3514	790\\
-2.2792	1000\\
-2.207	1212\\
-2.1348	1412\\
-2.0626	1592\\
-1.9904	1691\\
-1.9182	1739\\
-1.846	1827\\
-1.7738	1786\\
-1.7016	1800\\
-1.6294	1745\\
-1.5572	1724\\
-1.485	1638\\
-1.4128	1562\\
-1.3406	1468\\
-1.2684	1440\\
-1.1962	1345\\
-1.124	1301\\
-1.0518	1271\\
-0.9796	1239\\
-0.9074	1128\\
-0.8352	1148\\
-0.763	1132\\
-0.6908	1126\\
-0.6186	1109\\
-0.546400000000001	1094\\
-0.474200000000001	1108\\
-0.402000000000001	1138\\
-0.329800000000001	1126\\
-0.2576	1173\\
-0.1854	1223\\
-0.1132	1237\\
-0.0410000000000004	1221\\
0.0311999999999997	1101\\
0.1034	781\\
0.1756	184\\
0.247799999999999	3\\
0.319999999999999	0\\
0.392199999999999	0\\
0.464399999999999	0\\
0.5366	0\\
0.6088	0\\
0.681	0\\
0.753199999999999	0\\
0.8254	0\\
0.897599999999999	0\\
0.9698	0\\
1.042	0\\
1.1142	0\\
1.1864	0\\
1.2586	3\\
1.3308	8\\
1.403	22\\
1.4752	23\\
1.5474	51\\
1.6196	60\\
1.6918	73\\
1.764	71\\
1.8362	61\\
1.9084	57\\
1.9806	28\\
2.0528	20\\
2.125	16\\
2.1972	5\\
2.2694	0\\
2.3416	1\\
2.4138	3\\
2.486	3\\
2.5582	14\\
2.6304	36\\
2.7026	42\\
2.7748	78\\
2.847	80\\
2.9192	85\\
2.9914	66\\
3.0636	41\\
3.1358	27\\
3.208	22\\
3.2802	9\\
3.3524	14\\
3.4246	52\\
3.4968	83\\
3.569	131\\
3.6412	121\\
3.7134	63\\
3.7856	27\\
3.8578	4\\
3.93	4\\
};
\addlegendentry{\footnotesize eigenvalues}

\addplot[ybar, bar width=1.5, fill=green, draw=black, area legend] table[row sep=crcr] {%
0.549539784816691	300\\
};
\addplot[forget plot, color=white!15!black] table[row sep=crcr] {%
-4.09	0\\
4.73	0\\
};
\addlegendentry{\footnotesize $\bar{\lambda}$ threshold}

\end{axis}
\end{tikzpicture}%
\end{tabular}
\caption{Eigenvalue distributions. Left: SCM of observations. Middle: Tyler covariance matrix of observations. Right: Tyler covariance matrix of observations after whitening process. K-distributed case with shape parameter $\nu=0.5$, $\rho=0.8$, $m=100$, $N=1000$ ($c=0.1$), $K=3$, $\log(\bar{\lambda})=\log(1.7325)$.}
\label{allfig}
\end{figure*}
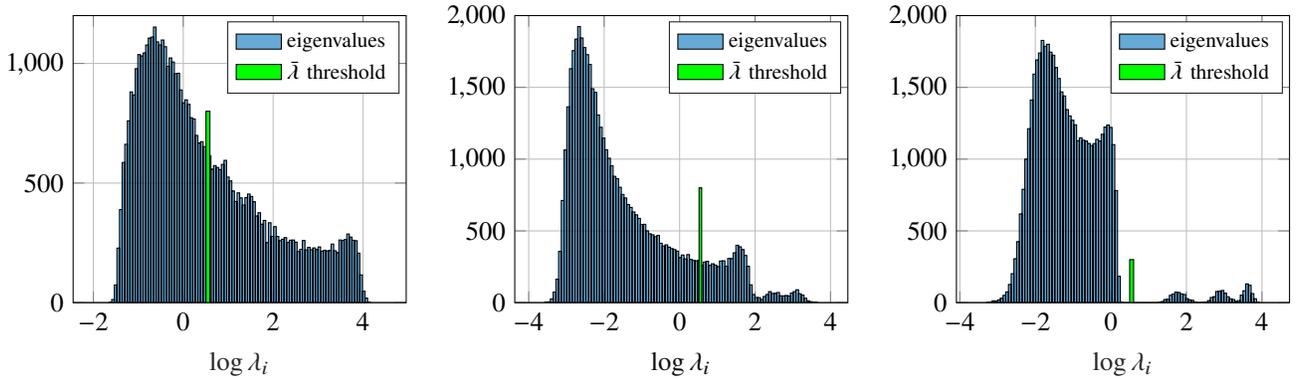

\textit{Notations}: Matrices are in bold and capital, vectors in bold. $Tr(\mathbf{X})$ is the trace of the matrix $\mathbf{X}$. $\left\Vert \mathbf{X} \right\Vert$ stands for the spectral norm. For any matrix $\mathbf{A}$, $\mathbf{A}^T$ is the transpose of $\mathbf{A}$.
 For any $m-$vector $\mathbf{x}$, $\mathcal{L}: \, \mathbf{x} \mapsto  \mathcal{L}(\mathbf{x})$ is the $m \times m$ matrix defined as the Toeplitz operator: $\left(  [\mathcal{L} (\mathbf{x})]_{i,j} \right) _{i\leq j} = x_{i-j}$ and $\left( [\mathcal{L} (\mathbf{x})]_{i,j} \right)_{i>j} = x_{i-j}^\ast$. For any matrix $\mathbf{A}$ of size $m \times m$, $\mathcal{T}(\mathbf{A})$ represents the matrix $\mathcal{L}(\check{\mathbf{a}})$ where $\check{\mathbf{a}}$ is a vector for which each component  $\check{\mathbf{a}}_{i ,  \, 0<i<m-1}$ contains the sum  of the $i-$th diagonal of $\mathbf{A}$ divided by $m$. 
 \vspace{-0.2cm}

\section{Model and assumptions}
Suppose that our investment universe is composed of $m$ assets characterized at each time $t$ by their returns. Let's denote by $\mathbf{R} = [\mathbf{r}_1, \cdots, \mathbf{r}_N]$ the $m \times N$-matrix containing $N$ observations (or return $m$-vectors) $\{\mathbf{r}_t\}_{t \in [1,N]}$ at date $t$. We assume next that the returns of the $m$ assets can conjointly be expressed as a multi-factor model where an unknown number $K < m$ of factors may be characteristic of this universe (i.e. among the $m$ assets, there exists $K$ principal factors that are driving the universe comprising these particular $m$ assets). We assume the additive noise to be a multivariate Elliptical Symmetric noise \cite{Kelker70,Ollila12} generalizing a correlated multivariate non-Gaussian noise. We then have, for all $t \in[1, N]$: $\mathbf{r}_t = \mathbf{B}_t \, \mathbf{f}_t  + \sqrt{\tau_t} \, \mathbf{C}^{1/2}\, \mathbf{x}_t$ where
\begin{itemize}
\item[$\bullet$] $\mathbf{r}_t$ is the $m$-vector of returns at time $t$, 
\item[$\bullet$] $\mathbf{B}_t$ is the $m \times K$-matrix of coefficients that define the sensitivity of the assets to each of the factor at time $t$,
\item[$\bullet$] $\mathbf{f}_t$ is the K-vector of factor values at t, supposed to be common to all the assets,
\item[$\bullet$] $\bdm{x}_t$ is a zero-mean unitarily invariant random $m$-vector of norm  $||\bdm{x}_t||^2=1$, 
\item[$\bullet$] $\bdm{C}$ is called the $m \times m$  scatter matrix (equal to the covariance matrix up to a constant) and is supposed to be Toeplitz structured and time invariant over the period of observation,
\item[$\bullet$] $\tau_t$ is a real positive random variable at $t$ representing the variance of the noise. This quantity is different along the time $t$ and can efficiently pilot the non-Gaussian nature of the noise.   
\end{itemize}
\noindent The efficient estimation of the number of factors $K$ is really a challenging problem for many financial applications:
\begin{itemize}
\item[$\bullet$] identifiability of the main $K$ factors to build new portfolios. This problem is for example closely related to linear unmixing problem in Hyperspectral Imaging \cite{Bioucas12},
\item[$\bullet$] identifiability of the main $K$ factors to separate signal and noise subspaces in order to build projectors, to filter noisy part of the data  through  jointly  robust and efficient covariance matrix estimation. This is for example useful for portfolio allocation or in risk management \cite{melas09,jay11, jayduvdar13, dargourjayGI}. 
\end{itemize}
The identified theoretical problem to solve is clearly the model order selection estimation  as well as efficient method of covariance matrix estimation under correlated non-Gaussian noise hypothesis.

\section{Maximum Variety Portfolio}
Portfolio allocation is a widely studied problem. Depending on the investment objective, the portfolio allocation differs. Apart from the well-known methods resides the differentiating Maximum Variety process that aims at maximising the Variety Ratio of the final portfolio.
One way to quantify the degree of diversification of a portfolio invested in $m$ assets with proportions $\bdm{w}=[w_1, \ldots, w_m]^T$ is to compute the Variety Ratio of the portfolio: \vspace{-0.1cm}
\begin{equation}\label{eq_varietyRatio}
\displaystyle 
VR(\bdm{w}, \bds{\Sigma}) = \frac{\bdm{w}^T \, \bdm{s}}{\left(\bdm{w}^T \, \bds{\Sigma} \, \bdm{w}\right)^{1/2}} \, ,\vspace{-0.1cm}
\end{equation}
\noindent where $\bdm{w}$ is the $m$-vector of weights, $w_i$ representing the allocation in asset $i$, $\bds{\Sigma}$ is the $m \times m$ covariance matrix of the $m$ assets returns and where $\bdm{s}$ is the $m$-vector of the square roots of the diagonal element of $\bds{\Sigma}$, ie $s_i = \sqrt{\bds{\Sigma}_{ii}}$, representing the standard deviation of the returns of the $m$ assets.
One way to allocate among the assets would be to maximise the above diversification ratio with respect to the weight vector $\bdm{w}$ to obtain the solution $\bdm{w}^*_{vr}$, also called the Maximum Diversified Portfolio in \cite{choueifaty08}: \vspace{-0.1cm}
\begin{equation}\label{eq_maxVR}
\displaystyle 
\bdm{w}^*_{vr} = \underset{\bdm{w}}{\mathrm{argmin}} \, VR(\bdm{w}, \bds{\Sigma}) \, ,\vspace{-0.1cm}
\end{equation}
\noindent under some conditions and constraints on the individual values of $\bdm{w}$. In the following, we will impose only $0 \leq w_i \leq 1$ $\forall i\in [1, m]$ and $\displaystyle\sum_{i=1}^m w_i=1$. 
As the objective function in (\ref{eq_maxVR}) depends on the unknown covariance matrix $\bds{\Sigma}$,  this latter has to be estimated in order to get the portfolio composition. This problem is one of the challenging problems in portfolio allocation and several methods can apply. The optimisation problem is shown to be very sensitive to outliers and to the chosen method of covariance matrix estimation. One of the main technique consists first in building a de-noised covariance matrix by thresholding the lowest eigenvalues and then in solving the objective function. The open questions always remain the choice of the covariance matrix estimate as well as the choice of the threshold value. To overcome these drawbacks and to answer these two questions, we propose a robust and quite simple technique based both on the class of the robust $M$-estimators and the RMT. 

\section{Proposed Methodology}
Under general non-Gaussian noise hypothesis proposed in Section II, Tyler $M$-estimator  \cite{Tyler87,Pascal8b} is shown to be the most robust covariance matrix estimate. Given $N$ observations of the $m$-vector $\bdm{r}_t$, the Tyler-$M$ estimate $\widehat{\bdm{C}}_{tyl}$ is defined as  the solution of the following "fixed-point" equation: \vspace{-0.1cm}
\begin{equation}\label{eq_tylerM}
\bdm{C} = \displaystyle \frac{m}{N} \sum_{t=1}^{N} \frac{\bdm{r}_t \, \bdm{r}^T_t}{\bdm{r}^T_t \, \bdm{C}^{-1}\, \bdm{r}_t}, \vspace{-0.1cm} 
\end{equation}
\noindent with $Tr(\widehat{\bdm{C}}_{tyl})=m$. The scatter matrix, solution of (\ref{eq_tylerM}) has some remarquable properties \cite{Pascal8,Mahot12} like being robust and "variance"-free and really reflects the true structure of the underlying process without power pollution. When the sources are present in the observations $\left\{\mathbf{r}_t\right\}$, the use of this estimator may lead to whiten the observations and to destroy the main information concentrated in the $K$ factors. \\
\indent When the noise is assumed white distributed, several methods, based on the RMT have been proposed \cite{Couillet11} to extract information of interest from the received signals. One can cite for instance the number of embedded sources estimation \cite{Kritchman9}, the problem of radar detection \cite{7383758}, signal subspace estimation \cite{Hachem13}. However, when the additive  noise is  correlated, some RMT methods require the estimation of a specific threshold which has no explicit expression and can be very difficult to obtain \cite{Vinogradova13, Couillet15b} while the others  assume that the covariance matrix is known and use it, through some source-free secondary data, to whiten the signal. According to the following {\it consistency theorem} found and proved in \cite{terreaux17, terreaux17a, terreaux18}, recent works have proposed to solve the problem through a biased Toeplitz estimate of $\widehat{\bdm{C}}_{tyl}$, let's say $\widetilde{\bdm{C}}_{tyl}=\mathcal{T}\left(\widehat{\bdm{C}}_{tyl}\right)$:
\begin{theorem}
Under the RMT regime assumption, ie that $N, m \rightarrow \infty$, and the ratio $c = m/N \rightarrow c > 0$, we have the following spectral convergence:
\begin{equation}\label{eq_consistency}
\left\Vert \mathcal{T}\left(\widehat{\bdm{C}}_{tyl}\right) - \mathbf{C} \right\Vert \stackrel{a.s.}{\longrightarrow} 0.
\end{equation}
\end{theorem}
This powerful theorem says that it is possible to estimate the covariance matrix of the correlated noise even if the observations contain the sources or information to be retrieved. According to this result, the first step is then to whiten the observations using $\widetilde{\bdm{C}}_{tyl}$. The whitened observations are defined as $\bdm{r}_{w,t} = \tildeC_{tyl}^{-1/2} \, \bdm{r}_t$. \\
\indent Given the set of $N$ whitened observations $\left\{\bdm{r}_{w,t}\right\}$ and given the Tyler's covariance matrix $\hat{\bds{\Sigma}}_w$ of these whitened returns, recent work \cite{terreaux18} has shown that this whitening process allows us to consider that the eigenvalues distribution of $\bds{\Sigma}_w$ has to fit the predicted bounded distribution of Mar\v{c}enko-Pastur \cite{Marchenko67} except for a finite number of eigenvalues if any source is still present and powerful enough to be detected outside the upper bound of the Mar\v{c}enko-Pastur distribution given by  $\bar{\lambda} = \left(1+\sqrt{c}\right)^2$.\\
\indent Figure \ref{allfig} compares the eigenvalues distribution of the SCM $\hat{\mathbf{C}}_{scm} = \mathbf{R} \, \mathbf{R}^T /N$,  $\hat{\mathbf{C}}_{tyl}$ and $\hat{\bds{\Sigma}}_w$ for $K=3$ sources of information embedded in non-Gaussian correlated K-distributed noise. If no whitening operation is made before applying the Mar\v{c}enko-Pastur boundary properties of the eigenvalues, then there is no chance to detect any of the sources. After whitening process, the only detected sources above the Mar\v{c}enko-Pastur threshold correspond to the $K$ sources. As a matter of fact, there is no need anymore to adapt the value of the threshold value regarding the distribution of $\tau_t$ and the estimated value of $\Bbb{E}[\tau]$ \cite{terreaux18}. The robust Tyler $M$-estimator is "$\tau$-free", i.e. it does not depend anymore of the distribution of $\tau_t$.
Once the $K$ largest eigenvalues larger than $\bar{\lambda}$ are detected, we set the $m-K$ lowest ones to $\left(Tr\left(\hat{\bds{\Sigma}}_w\right)-\sum_{k=K+1}^{m}\lambda_k\right) / (m-K)$, and then build back the de-noised covariance matrix to be used in \eqref{eq_maxVR} (or in any other objective function).

\section{Application}
\begin{figure*}
\centering
\includegraphics[width=0.7\columnwidth]{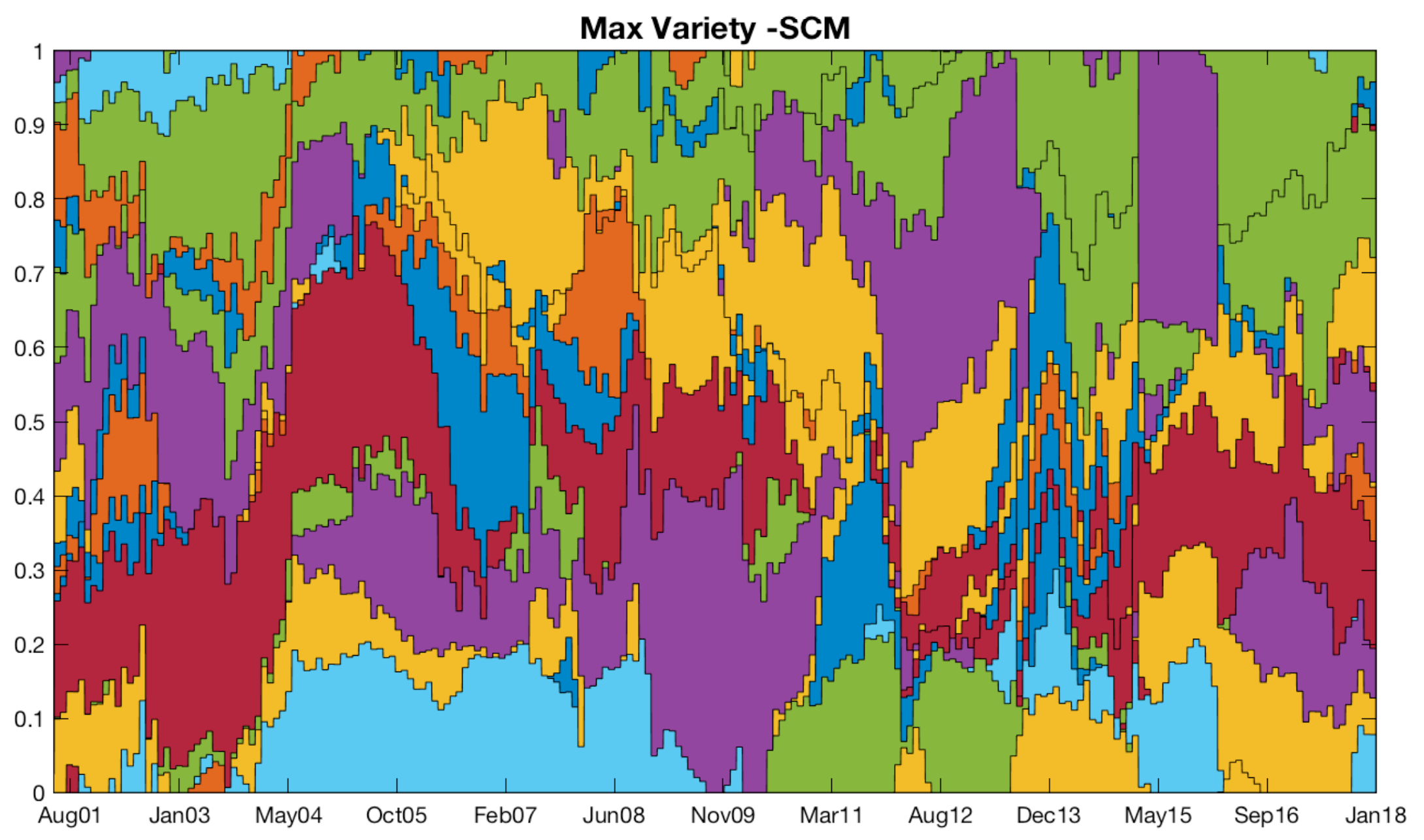} 
\includegraphics[width=0.7\columnwidth]{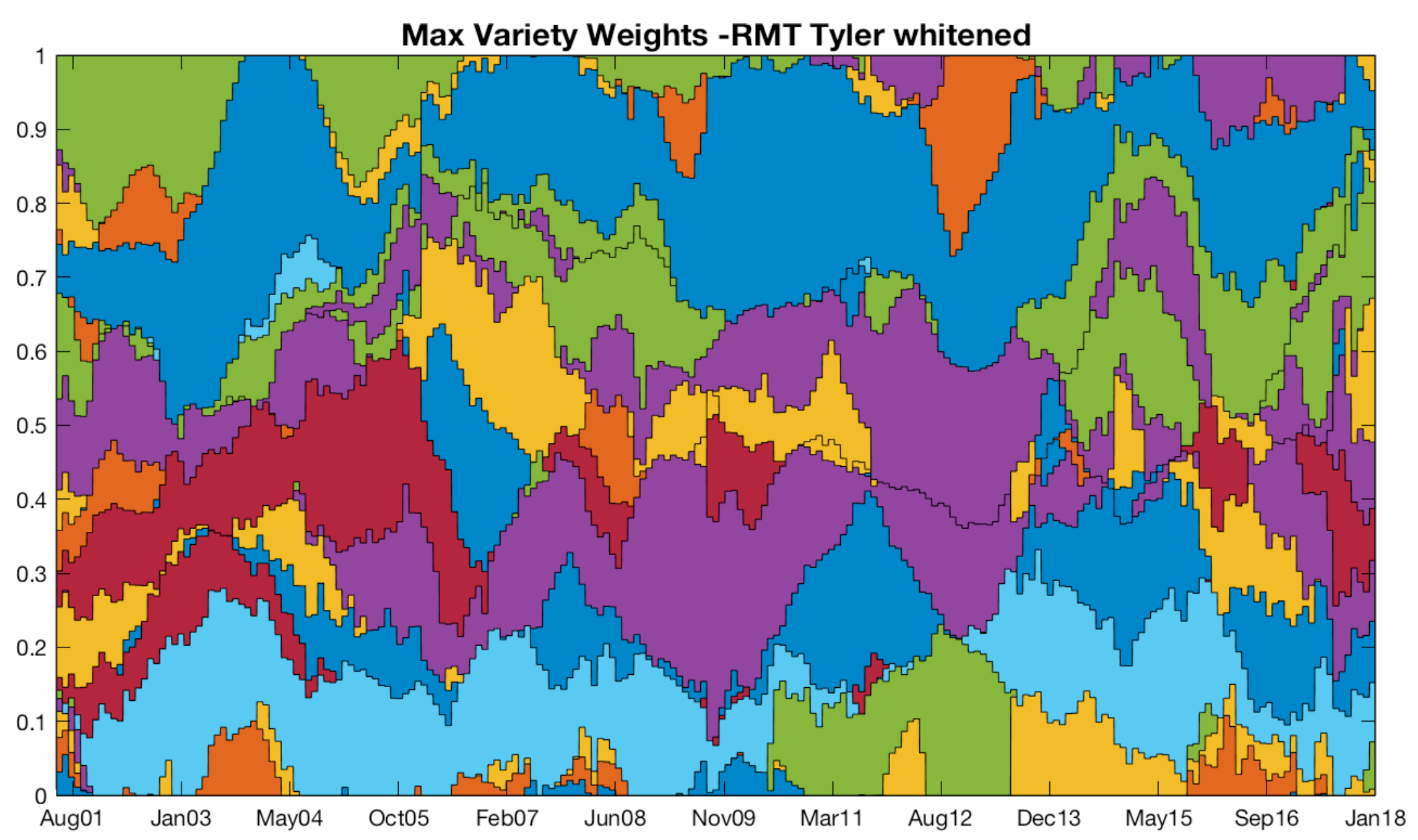}
\includegraphics[width=0.5\columnwidth]{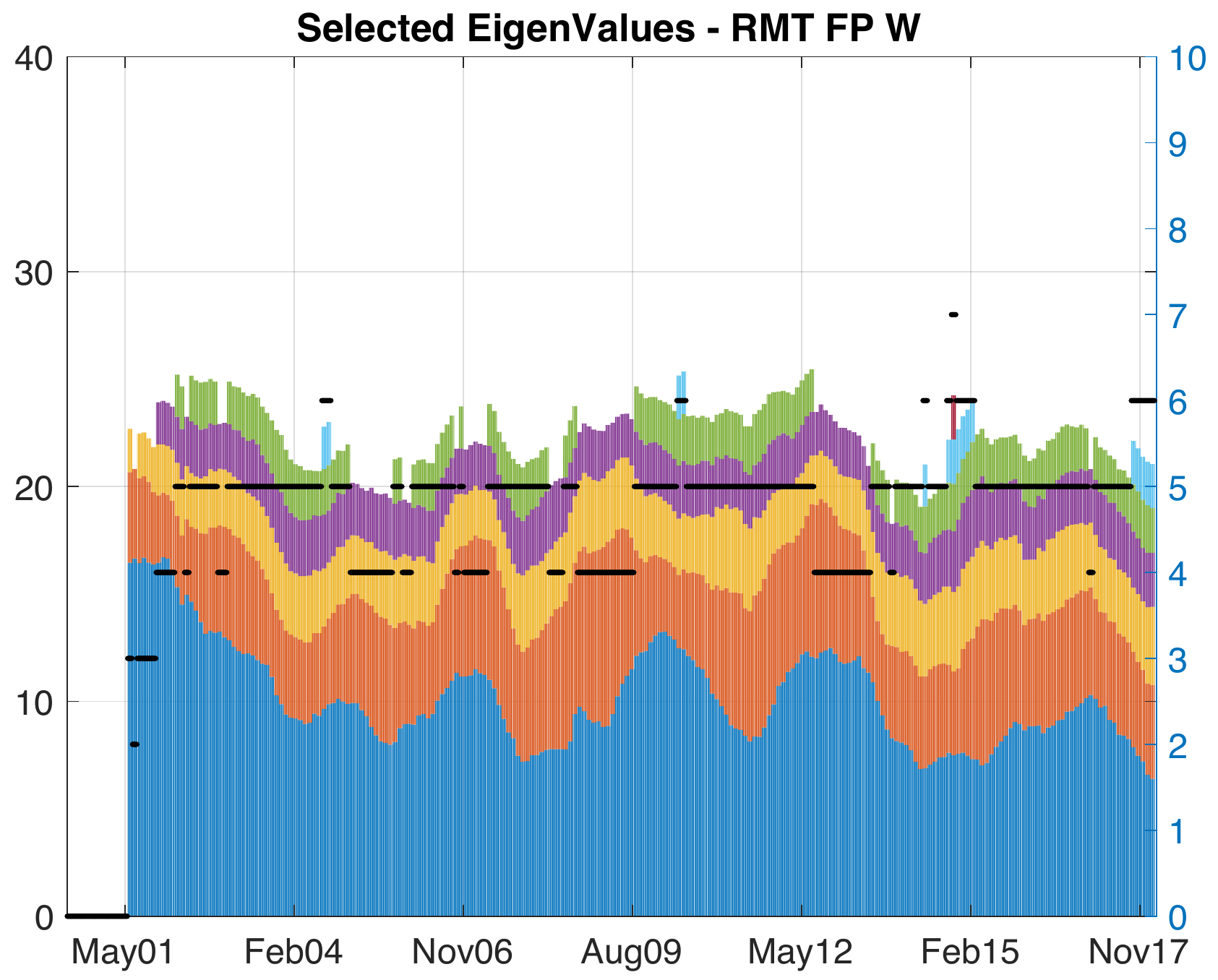}
\caption{Left and middle: dynamic weights as a stacked area chart. Each colour represents an asset. The Variety Max "RMT Tyler whitened" weights change smoother than the ones obtained with SCM, confirmed also by a smaller cumulative turnover. Right: values of the selected eigenvalues (left axis) and their number (right axis).}.
\label{fig_data_weights}
\end{figure*}

This section is devoted to show the improvement of such a process when applied to the Maximum Variety Portfolio process. This allocation process (denoted as "Variety Max" in the following) is the one designed and used by Fideas Capital for allocating their portfolios. The investment universe consists of $m=40$ baskets of European equity stocks representing twenty-one industry subsectors (e.g. transportation, materials, media...), thirteen countries (e.g. Sweden, France, Netherlands,...) and six factor-based indices (e.g. momentum, quality, growth, ...). Using baskets instead of single stocks allows to reduce the idiosyncratic risks and the number of assets to be considered. We observe the prices of these assets on a daily basis from June 2000, the 19th to January 2018 the 29th. The daily prices are close prices, i.e. the price being fixed before the financial marketplaces close at the end of each weekday. 
The portfolios weights are computed as follows: every four weeks, we estimate the covariance matrix of the assets using the past one year of returns and we run the optimisation procedure in order to get the vector of weights that maximises the variety ratio (\ref{eq_varietyRatio}) given this past period. The computed weights, say at time $t$, are then kept fixed for the next four weeks period. We compare the results obtained with the proposed methodology with the ones obtained using the SCM and we report several numbers in order to compare the benefits of such a method. Performance are also compared to the STOXX$^{\mbox{\scriptsize{\textregistered}}}$ Europe $600$ Index \cite{stoxx600} performance that is composed of $600$ large, mid and small equity stocks across $17$ countries of the European regions.
\begin{figure*}
\centering\includegraphics[width=1.33\columnwidth]{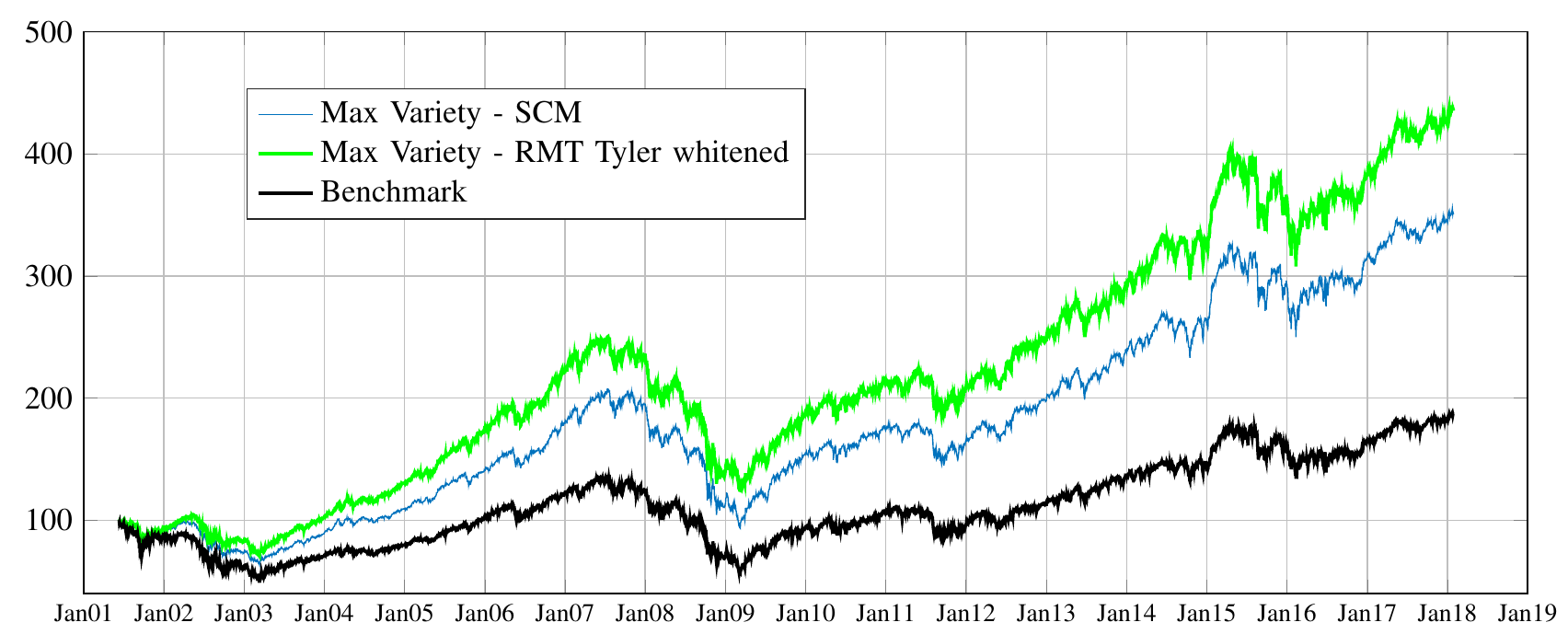}
\includegraphics[width=0.67\columnwidth]{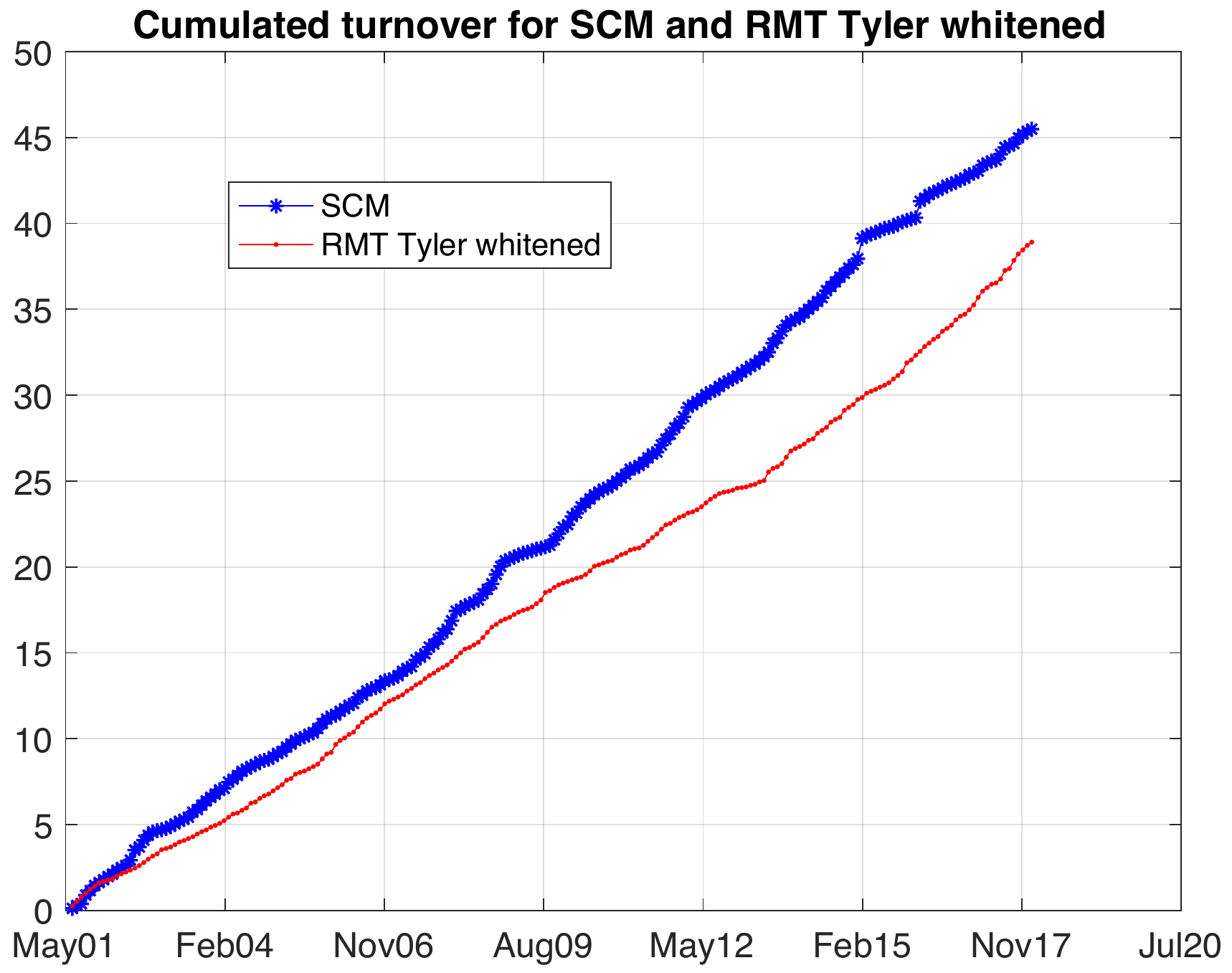}
\caption{Left: portfolios wealth starting at $100$ at the first period. Right: cumulative sum of absolute weight changes (turnover) between the consecutive periods.}
\label{fig_resPrice}
\end{figure*}
On the left of Figure \ref{fig_resPrice}, we report the evolution of portfolios wealths, starting at $100$ at the beginning of the first period. The Variety Max "SCM" and "RMT Tyler whitened" portfolios are respectively in blue and green and the price of the benchmark is the black line. The proposed RMT Tyler whitened technique clearly outperforms conventional ones. On the right of this figure, the cumulative turnover is shown for the both portfolios. We assume that the turnover (or the change in weights) between two consecutive periods $t$ and $t+1$ is measured by $\sum_{i=1}^m |w_{i,t+1}-w_{i,t}|$. Again, the proposed technique leads to lower the cumulated turnover which is important in finance. Limiting the turnover is often added as an additional non linear constraint to the optimization process \eqref{eq_maxVR}.\\
\indent Figure \ref{fig_data_weights} shows two different results. The two graphs on the left represent the evolution of the weights, on the overall period. Each colour represents an asset and the weights are stacked at each time (with the sum equals to one). The evolution of the weights for the Variety Max "RMT Tyler whitened" portfolio is smoother than for the SCM. This is confirmed by a lower turnover too, so that the allocation process is more stable when using the proposed methodology. On the right of the same figure, we report the values of the selected eigenvalues (on the left axis) and its number as well (on the right axis). Most of the time, five eigenvalues are detected. This results show a different picture than the general one where only one source (the "market") is outside the Mar\v{c}enko-Pastur bound. As noticed before, we get the same improvements as with other allocation process such as the Global Minimum Variance Portfolio.
\vspace{-0.2cm}
\begin{table}[h!]
\caption{Some performance numbers.}
\begin{center}
\begin{tabular}{r|c|c|c|c|}
Variety Max &	Ann. 			&	Ann.			& 	Ratio			& Max \\
Portfolios &	Return 		&	Volatility 		& 	(Ret / Vol)	& DD \\
\hline
\hline
RMT Tyler Whithened 	& 9,71\% 	& 12,9\% 	& 0,75	& 50,41\%\\
\hline
SCM				& 8,51\%		& 13,80\%	& 0,62	 & 55,02\%\\
\hline
Benchmark					& 4,92\%	 	& 15,19\% 	& 0,32	& 58,36\%\\
\hline 
\end{tabular}
\end{center}
\label{tab_stats}
\end{table}%
\vspace{-0.2cm}
We finally report on table \ref{tab_stats} some statistics on the overall portfolio performance: we compare, for the whole period, the annualised return, the annualised volatility, the ratio between the return and the volatility and the maximum drawdown of the portfolios and the benchmark. All the qualitative indicators related to the proposed technique show a significant improvement.
\vspace{-0.2cm}
\vspace{-0.2cm}


\section{conclusion}
In this paper we have shown that when processed correctly the Maximum Variety Portfolio allocation process leads to improved performance with respect to a classical approach. The improvement comes especially from the robust and denoised version of the covariance matrix estimate. Indeed, we have modelled the assets returns as a multi-factor model embedded in a correlated elliptical and symmetric noise, allowing to account for non-Gaussian and non correlated noise. Given this model setup, then we show how to separate the signal from the noise subspace using a "toeplitzified" robust and consistent Tyler-M estimator and the Random Matrix theory applied on the whitened covariance matrix estimate. This paper has taken the Maximum Variety Portfolio process as an example but the same results apply on other allocation framework involving covariance matrix estimation (and/or model order selection), such as the Global Minimum Variance Portfolio. Moreover this can also be exploited to define the main directions of information and to construct pure factor driven models. These methods have also shown their importance in the radar and hyperspectral fields and are very promising techniques for many applications.

\vspace{-0.2cm}
\section*{Acknowledgments}
\addcontentsline{toc}{section}{Acknowledgments}
We would like to thank DGA and Fideas Capital for supporting this research and providing the data. We thank particularly Thibault Soler, and also Pierre Filippi and Alexis Merville for their constant interaction with the research team at Fideas Capital.

\balance
\vspace{-0.2cm}
\bibliographystyle{IEEEtran}
\bibliography{biblio_ejay}

\end{document}